\documentclass{jps-cp}
\usepackage{txfonts} %Please comment out this line unless the txfonts package is availabe in your LaTeX system.

\newcommand{\mc}[1]{\mathcal{#1}}
\newcommand{\bra}[1]{\langle{#1}|}
\newcommand{\ket}[1]{|{#1}\rangle}

\title{Deuteron helicity flip generalized parton distributions in a convolution model}

\author{W. Cosyn$^{1}$ and B. Pire$^{2}$}

\inst{$^{1}$Department of Physics and Astronomy, Ghent University, 
             Proeftuinstraat 86, B9000 Ghent, Belgium \\
$^{2}$Centre de Physique Th\'{e}orique, \'{E}cole Polytechnique, CNRS, 
91128 Palaiseau, France}

\email{wim.cosyn@ugent.be}

\recdate{\today}

\abst{We discuss the general properties of generalized parton distributions with helicity flip (transversity) for spin-1 hadrons in the leading twist case.  Using a basic light cone convolution model, we show the deuteron helicity amplitudes containing quark helicity flip GPDs and comment on the role deuteron angular momentum plays in these.}

\kword{generalized parton distributions, deuteron, transversity}

\begin{document}
\maketitle

\section{Helicity flip GPDs for spin 1 hadrons}

Generalized parton distributions (GPDs) appear as scalar functions in the decomposition of off-forward quark and gluon correlators in hadrons.  Through QCD factorization theorems they parametrize the non-perturbative part of the amplitude in processes such as deeply virtual Compton scattering (DVCS) and deep exclusive meson production (DEMP)~\cite{Diehl:2003ny}.
While a rich phenomenology exists for the nucleon, there is comparatively less material available concerning nuclear GPDs, which enter in coherent exclusive processes on nuclei.  In this proceedings, we report on recent work on quark helicity flip GPDs for the deuteron.  More details can be found in Refs.~\cite{Cosyn:2018rdm,Cosyn:2018thq}.

As the deuteron is a spin 1 object, it admits more GPDs than the spin 1/2 case.  In the leading twist case, both for quark and gluons there are 9 helicity conserving GPDs and 9 helicity flip or transversity ones.  The helicity conserving ones were introduced in Ref.~\cite{Berger:2001zb}, while the helicity flip ones were recently introduced in Ref.~\cite{Cosyn:2018rdm}.  Both sectors evolve separately under QCD evolution, and for the helicity flip sector quarks and gluon operators do not mix.

For the helicity flip sector, the decomposition of the quark correlator for a spin-1 hadron takes the following form~\cite{Cosyn:2018rdm}
\begin{align} \label{eq:HTq_decomp}
T^{q\;i}_{\lambda' \lambda} = \frac{1}{2} &\int 
\frac{d\kappa}{2\pi}e^{ix\kappa(Pn)}\bra{p'\,\lambda'}\bar{\psi}(-\frac{\kappa}{2} 
n)(i n_\mu\sigma^{\mu i})\psi(\frac{\kappa}{2} n)\ket{p\,\lambda}=
M\frac{(\epsilon'^{*}n)\epsilon^i-\epsilon'^{*i}(\epsilon n)} {2\sqrt{2}(Pn)}
H^{qT}_1(x,\xi,t)\nonumber\\
%%%%%%%%%%%%%%%
&\qquad+M\left[\frac{2P^i(\epsilon n)(\epsilon'^{*} 
n)}{2\sqrt{2}(Pn)^2}-\frac{(\epsilon 
n)\epsilon'^{i*}+\epsilon^i(\epsilon'^{*} n)}{2\sqrt{2}(Pn)}\right] 
H^{qT}_2(x,\xi,t)\nonumber\\
%&&&&&&&&&&&&&&&&
&\qquad+\left[ \frac{(\epsilon'^{*} n)\Delta^i-\epsilon'^{i*}(\Delta n)}{2M(Pn)}(\epsilon 
\Delta) + \frac{(\epsilon n)\Delta^i-\epsilon^i (\Delta n)}{2M(Pn)}(\epsilon'^{*}\Delta)\right]  H^{qT}_3(x,\xi,t)\nonumber\\
%%%%%%%
&\qquad+\left[ \frac{(\epsilon'^{*} n)\Delta^i-\epsilon'^{i*}(\Delta n)}{2M(Pn)}(\epsilon 
\Delta) - \frac{(\epsilon n)\Delta^i-\epsilon^i (\Delta n)}{2M(Pn)}(\epsilon'^{*}\Delta)\right]  H^{qT}_4(x,\xi,t)\nonumber\\
%%%%%%%%%%%%%%%
&\qquad+ M\left[ \frac{(\epsilon'^{*} n)\Delta^i-\epsilon'^{i*}(\Delta n)}{2\sqrt{2}(Pn)^2}(\epsilon 
n) + \frac{(\epsilon n)\Delta^i-\epsilon^i (\Delta n)}{2\sqrt{2}(Pn)^2}(\epsilon'^{*}n)\right] 
H^{qT}_5(x,\xi,t)
\nonumber\\
%%%%%%%
&\qquad+ \frac{(\Delta^i+2\xi P^i)}{M}(\epsilon'^{*} \epsilon)H^{qT}_6(x,\xi,t)
-\frac{(\Delta^i+2\xi P^i)}{M} \frac{(\epsilon'^{*} \Delta)(\epsilon \Delta)}{4 M^2}
H^{qT}_7(x,\xi,t)\nonumber\\
%%%%%%%%%
&\qquad + \left[ \frac{(\epsilon'^{*} n)P^i-\epsilon'^{i*}(Pn)}{M(Pn)}(\epsilon 
\Delta) - \frac{(\epsilon n)P^i-\epsilon^i (Pn)}{2M(Pn)}(\epsilon'^{*}\Delta)\right] 
H^{qT}_8(x,\xi,t)\nonumber\\
%%%%%%%%%%%%
&\qquad + \left[ \frac{(\epsilon'^{*} n)P^i-\epsilon'^{i*}(Pn)}{M(Pn)}(\epsilon 
\Delta) + \frac{(\epsilon n)P^i-\epsilon^i (Pn)}{2M(Pn)}(\epsilon'^{*}\Delta)\right] 
H^{qT}_9(x,\xi,t)\,.
\end{align}
Here, $n$ is a light-like fourvector and $i$ a transverse index.  The initial spin-1 hadron (mass $M$) has fourvector $p$, polarization vector $\epsilon$ and light-front helicity $\lambda$, with the equivalent primed variables for the final hadron.  Kinematic variables are defined as follows
\begin{align}
&P=\frac{p'+p}{2}, &\Delta=p'-p, &&t=\Delta^2, &&\xi=-\frac{(\Delta n)}{2(Pn)}\,.
\end{align}
The 9 real GPDs have the following properties: 
\begin{enumerate}
\item 6 GPDs are even functions in skewness $\xi$ ($H^{qT}_i\,, i\in\{1,4,5,6,7,9\}$), the other 3 are odd in $\xi$.

\item Only one GPD does not decouple in the forward limit and can be linked to the collinear transversity distribution: $H^{qT}_1(x,0,0)=h_1(x)$.

\item General Mellin moments of the GPDs [where the $n$-th moment corresponds to an integral over $\int \mathrm{d}x \, x^{n-1} H^{Tq}_i(x,\xi,t)\,$] obey polynomiality sum rules, where the moments can be written as a polynomial in $\xi$ with generalized form factors in their coefficients.  For the $n$-th moment, there are $5+3\left(\lfloor \frac{n}{2}\rfloor \right)$ independent generalized form factors in the quark helicity flip sector~\cite{Cosyn:2018thq}.  Four transversity GPDs ($i\in\{2,3,5,8\}$) have zero sum rules for the first Mellin moment.
\end{enumerate}

\section{Deuteron convolution model}
To compute the quark transversity GPDs for the deuteron, we consider a basic convolution model.  We only consider the dominant $np$ component of the deuteron wave function and consider the leading order impulse approximation, where one of the nucleons acts as a so-called ``spectator''.  Using methods of light-front perturbation theory, nuclear and nucleon structure can be separated and we can write the correlator of Eq.~(\ref{eq:HTq_decomp}) as a convolution of two light-front deuteron wave functions and a quark helicity flip correlator for the nucleon~\cite{Cano:2003ju,Cosyn:2018rdm}.  The nucleon correlators are then decomposed through their corresponding spin 1/2 GPDs.

One disadvantage of only considering the lowest Fock state in the deuteron ($np$ component) is that this truncation breaks Lorentz covariance and consequently also the polynomiality requirements of the deuteron GPDs~\cite{Cosyn:2018rdm}.  Extensions including additional contributions to restore the polynomiality condition will be the topic of a future study.  

\begin{figure}[tbh]
\includegraphics[width=\textwidth]{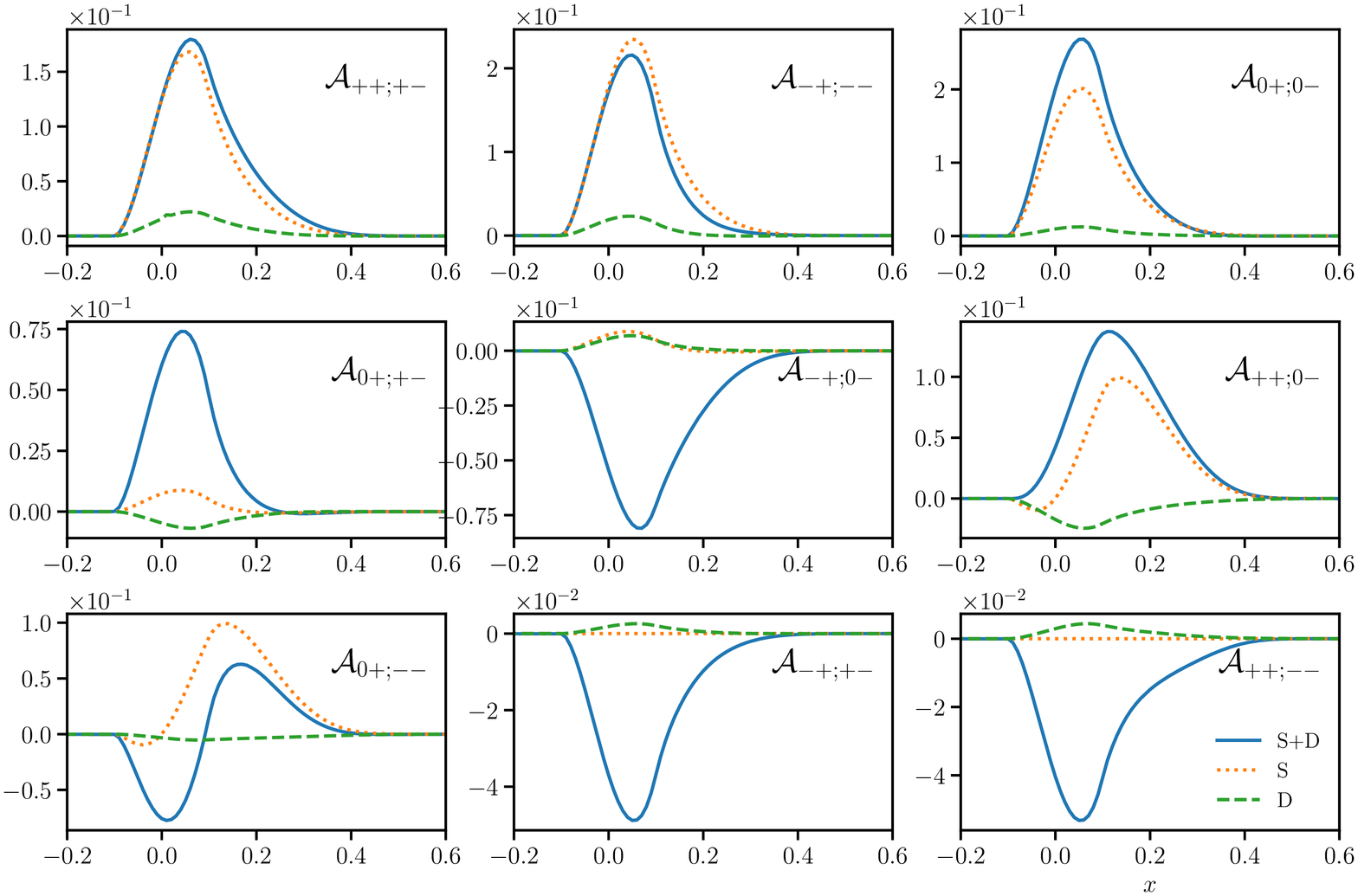}
\caption{Deuteron quark helicity amplitudes computed in the convolution formalism, at 
$\xi=0.1$,$t=-0.25$~GeV$^2$.  Full blue curve includes the full deuteron wave 
function, and dotted orange (dashed green) only includes the deuteron 
radial $S$-($D$-)wave.}
\label{fig:amp}
\end{figure}

\begin{figure}[tbh]
\includegraphics[width=\textwidth]{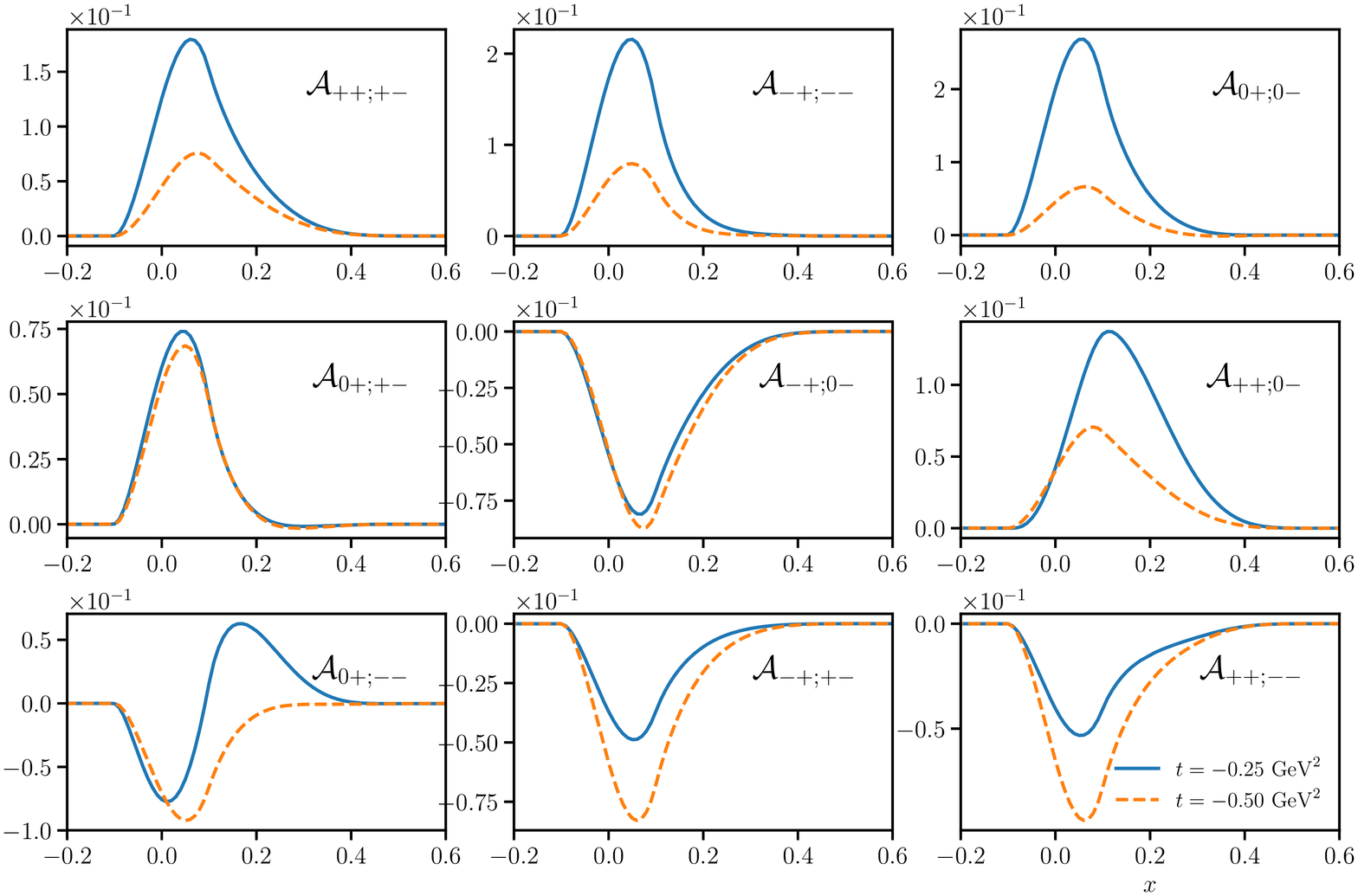}
\caption{Deuteron quark helicity amplitudes computed in the convolution formalism, at 
$\xi=0.1$ and two values of momentum transfer $t$.}
\label{fig:amp_t}
\end{figure}

In Figs.~\ref{fig:amp} and \ref{fig:amp_t} we show results in this convolution model for the helicity amplitudes $\mc{A}_{\lambda'+; \lambda - }$, where the plus and minus refer to the helicities of the outgoing and incoming quark in the correlator.  These helicity amplitudes form linear combinations of the transversity GPDs through the relation
\begin{equation}
\mc{A}_{\lambda'+; \lambda_d - } = \frac{1}{2} \left(T^{q1}_{\lambda' \lambda} + i \; T^{q2}_{\lambda' \lambda} \right) \,,
\end{equation}
and exhibit the role deuteron angular momentum in a more transparent way than the corresponding GPDs.  For plots of the GPDs, we refer to Ref.~\cite{Cosyn:2018rdm}.  For the calculations in the convolution model, we use the nucleon transversity GPDs of Ref.~\cite{Goloskokov:2011rd}, and the AV18 deuteron wave function parametrization~\cite{Wiringa:1994wb}.

Figure~\ref{fig:amp} shows the different contributions from the deuteron S- and D-wave 
components to the total result.  The difference between the first two and the latter originates from S-D interference contributions.  Considering the top row of Fig.~\ref{fig:amp}, which correspond to the deuteron helicity conserving amplitudes, it is clear that these are dominated by the pure S-wave contribution, whereas the other amplitudes that admit a change in deuteron helicity receive major contributions from the S-D interference terms.  The two amplitudes with two units of deuteron helicity flip (bottom row, right two panels) are identically zero when only including the deuteron S-wave as in that case there is no orbital angular momentum available in the deuteron to compensate the change in helicities (two units for the deuteron, one for the quark).

Figure~\ref{fig:amp_t} shows the helicity amplitudes at two values of momentum transfer $t$.  Helicity amplitudes with different units of deuteron helicity change show different behavior with increasing momentum transfer: for no helicity flip the amplitude shrinks with larger $t$, the amplitudes with a single unit of helicity change increase a little bit in size at the larger $t$ value, and the amplitudes with a complete deuteron helicity flip grow significantly larger.  This again reflects the role deuteron angular momentum plays, supplied through the momentum transfer.

To conclude, the role of these GPDs for the deuteron could be explored in the phenomenology of coherent DVCS (where gluon transversity enters at NLO) on the deuteron, double vector meson production (See Refs.~\cite{Ivanov, Enberg} for the nucleon case) and DEMP (in combination with a higher twist distribution amplitude)~\cite{Ahmad, Goloskokov:2011rd} on deuteron targets.

\end{document}